# Cost-Effective Frequency Planning for Capacity Enhancement of Femtocellular Networks

Mostafa Zaman Chowdhury[1], Yeong Min Jang[1], and Zygmunt J. Haas[2]

[1]*Department of Electronics Engineering, Kookmin University, Seoul, South Korea*

[2] *Wireless Networks Lab, Cornell University, Ithaca, NY, U.S.A*

E-mail: mzceee@yahoo.com, yjang@kookmin.ac.kr, zhaas@cornell.edu

**Abstract:** The *femto-access-point*, a low-cost and small-size cellular base-station, is envisioned to be widely deployed in subscribers' homes, as to provide high data-rate communications with improved quality of service. As *femtocellular* networks will co-exist with *macrocellular* networks, mitigation of the interference between these two network types is a key challenge for successful integration of these two technologies. In particular, there are several interference mechanisms between the femtocellular and the macrocellular networks, and the effects of the resulting interference depend on the density of *femtocell*s and the overlaid *macrocell*s in a particular coverage area. While improper interference management can cause a significant reduction in the system capacity and can increase the outage probability, effective and efficient frequency allocation among femtocells and macrocells can result in a successful co-existence of these two technologies. Furthermore, highly dense femtocellular deployments – the ultimate goal of the femtocellular technology – will require significant degree of self-organization in lieu of manual configuration. In this paper, we present various femtocellular network deployment scenarios, and we propose a number of frequency-allocation schemes to mitigate the interference and to increases the spectral efficiency of the integrated network. These schemes include: *shared frequency band*, *dedicated frequency band*, *sub-frequency band*, *static frequency-reuse*, and *dynamic frequency-reuse*. We derive an analytical model, which allows us to analyze in details the user's outage probability, and we compare the performance of the proposed schemes using numerical analysis.

**Keywords:** Femtocell, Femtocellular Network, Overlay Networks, Interference Management, Interference Mitigation, Frequency Allocation, Self-Organizing Networks, SON, Outage Probability.

## 1. Introduction

With the proliferation of various multimedia traffic types, future wireless networks will necessitate high data-rates with improved quality of service (QoS) and at low cost. Small-size, inexpensive, and low-power *femto-access-point* (*FAP*) can support the growing demand for larger bandwidth and better QoS for the indoor communication needs. Indeed, a well-designed *femtocellular* network can divert large amounts of traffic from the congested and expensive *macrocellular* networks to femtocellular networks. One of the key advantage of the femtocellular technology is in the fact that it uses the same frequency bands as the macrocellular networks, thus avoiding the need to introduce new user equipment. However, the use of the same frequency spectrum can also cause substantial interference if no adequate interference management is incorporated into the network design. Indeed, network architecture, handover control, and interference management are key issues for cost-effective integrated femtocell/macrocell network deployment [1]. Network architecture and resource management depend mostly on the size of the femtocells deployment, the existing network infrastructure, and the future extension plan.

Interference between the two technologies could be managed through proper frequency allocation schemes, which would allow largest utilization of the valuable radio spectrum and the highest level of user's quality of experience (QoE). Specifically, appropriate interference management, implemented through suitable frequency-allocation schemes, increases the system capacity, reduces the outage probability, and increases the frequency utilization. And although the interference in a femtocellular network cannot be fully eliminated, it is possible to reduce the interference to within a reasonable range by proper management.

Interference mitigation using the same frequency allocation scheme [2-5] for the dense and the sparse femtocellular network deployments is not efficient. Rather, the choice of the frequency-allocation scheme to minimize the interference and to ensure maximal spectrum utilization should depend on the density of the femtocells and the relationship between the femtocells and the macrocells, which are the two parameters that we use for classification of the deployment scenarios. The different frequency-allocation schemes for different femtocellular network environments proposed in this paper increase the frequency spectrum utilization, providing excellent performance in terms of cost, capacity, and outage probability.

In the *shared frequency band* scheme, the same frequency band is allocated for all the femtocells and the macrocells, while in the *sub-frequency band* scheme, the whole frequency band is allocated for the macrocells and part of the whole frequency band is allocated for the femtocells. Both of these schemes can provide good performance only for a small-scale femtocellular network. In the *static frequency-reuse* scheme, which is proposed in this paper, two frequency bands are allocated to the femtocells within a macrocell and a different frequency band is allocated to the overlaid macrocell. Based on this scheme, almost half of the femtocells within a macrocell use one frequency band and the remaining half use the other another frequency band. This scheme can improve the performance of a non-dense femtocellular network, but not of a dense femtocellular network. The *static frequency-reuse* scheme does not utilize the *self-organizing network* (*SON*) feature.

For a dense femtocellular network, we propose the *dynamic frequency-reuse* scheme, where two different frequency bands are allocated for the femtocells. One frequency band is allocated for the inner area of a femtocell, and the other frequency band is allocated for the outer area of the femtocell. The *dynamic frequency-reuse* scheme uses the *SON* feature. The frequency assignment and the transmission power of femtocell users are automatically adjusted to mitigate the inter-femtocell interference. Using the *SON* feature, densely deployed *FAP*s perform self-configuration, self-optimization, and self-healing [6] of the frequency assignment and of power optimization to mitigate interference effect from other neighboring femtocells. The *dynamic reuse frequency* scheme is proposed here for the medium- and the highly-dense femtocellular network to increase the spectral efficiency and to mitigate the effects of interference.

In both, the *static* and the *dynamic frequency-reuse* schemes, the frequencies allocated for the femtocells within a macrocell and the frequencies allocated to the overlaid macrocell are distinct, so that the femtocell-macrocell interference is essentially eliminated. The total frequency band allocated for the femtocells within a particular macrocell in a macrocell cluster is the sum of the frequency bands of the other macrocells of the macrocell cluster.

Of course, the outage probability increases with the increase of interference within a femtocellular network. The outage probability calculations in this paper consider the interference from neighboring femtocells, from overlaid macrocell, and from neighboring macrocells.

The rest of this paper is organized as follows. Section 2 introduces a number of femtocellular network deployment scenarios and their corresponding interference patterns. Various frequency-allocation schemes for interference mitigation are proposed in Section 3. Section 4 discussed the femtocellular network architecture based on the concept of Self-Organizing Network (SON). The outage probability of a femtocellular user is analyzed in Section 5 and the performance numerical results of the proposed schemes are presented in Section 6. Finally, conclusions are drawn in Section 7.

## 2. Interference Scenarios in a Femtocellular Network

There are different scenarios in which interference, created due to the co-existence of macrocells and femtocells in the same geographical area, may affect the performance of the femtocellular network. The amount of interference depends on the network architecture, location of femtocells, and density of femtocells. Based on these factors, Fig. 1 depicts four scenarios for femtocellular deployment.

- *Scenario A (Single femtocell without overlaid macrocells):* In this case, there is no interference effect from other cells. This scenario is typical of remote areas and interference is not an important issue in this case.

- *Scenario B (Single stand-alone femtocells overlaid by a macrocell):* In this case, there is a single femtocells overlaid by macrocell, so there is no femtocell-to-femtocell interference. However, a significant amount of femtocell-to-macrocell interference could exist in this case.

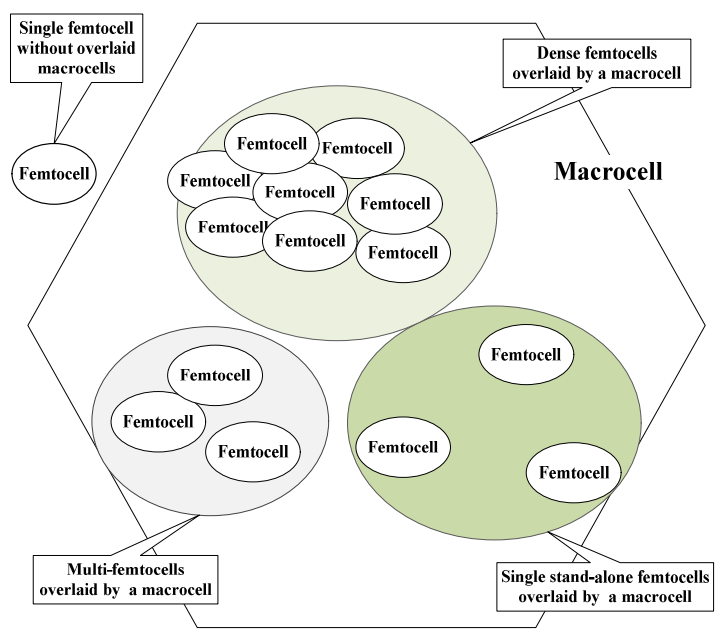

**Fig. 1:** Various scenarios for femtocellular/macrocellular interference

- ***Scenario C (Multi-femtocells overlaid by a macrocell):*** In this case, there are few neighboring femtocells in addition to the overlaid by macrocell. The transmissions of the four basic network entities: the macrocellular base station (BS), the macrocellular user equipment (UE), the femto-access-points (FAPs), and the femtocellular user equipment, all potentially affect each other.
- ***Scenario D (Dense femtocells overlaid by a macrocell):*** In this case, many femtocells are deployed in a relatively small geographical area. Although this is the scenario that is the goal of a successful femtocellular deployment, it also presents the worst case of interference. As in scenario C, the four basic network entities are all potentially affected by mutual interference.

Hence, the deployment scenarios *B* or *C* or the combination of deployment scenarios *B* and *C* are termed "non-dense femtocellular network deployment," while the deployment scenario *D* is referred to as "dense femtocellular network deployment."

The identity of the offenders (entities generating the interference) and the victims (entities affected by the interference) depend on the relative positions of the four basic network entities: the FAPs, the macrocellular UE, the femtocellular UE, and the macrocellular BS. The four different link types, the macrocellular downlink and uplink, and the femtocellular downlink and uplink, can potentially create harmful interference affecting the other basic network entities [7].

*Macrocell downlink:* The femto UEs within a macrocell coverage area receive interference from the macrocell downlink if both the macrocell and femtocell are allocated the same frequency. The situation is of particular concern when the location of the femtocell is close to the macrocellular BS and the femtocell UE is located at the edge of the femtocell, so that the transmitted power from the macrocellular BS can potentially cause severe interference to the femto UE receiver. This situation can occur in every one of the deployment scenarios *B, C, and D,* and is demonstrated in Fig. 2 by the macrocell downlink causing interference to the femto UE-1, femto UE-2, and femto UE-3.

*Macrocell uplink:* Whenever a macro UE is inside the femtocell coverage area or just close to a femtocell, the uplink signal from the macro UE to the macrocellular BS can cause interference to the FAP receiver. This situation can occur in the deployment scenarios *B, C,* and *D*. Fig. 2 shows an example where the macrocell uplink of macro UE-1 causes interference to FAP-2.

*Femtocell downlink:* In this case, the femtocell downlink causes interference to the macro UE receivers and to the nearby femto UE receivers. Whenever a macro user is inside or near a femtocell coverage area, the macro UE is subjected to interference from the femtocell downlink (deployment scenarios *B, C,* and *D*). Similarly, when two (or more) femtocells are in close proximity to each other, then the femto UE of one femtocell is affected by the interference from the neighbor femtocell downlink (scenarios *C* and *D*). Figure 2 depicts a situation where the downlink of the FAP-2 femtocell causes interference to macro UE-1 and femto UE-1.

*Femtocell uplink:* When a femtocell is close to the macrocellular BS, the transmitted uplink signal from the femto UE causes interference to the macrocell receivers (deployment scenarios *B, C,* and *D*). Similarly, when two (or more) femtocells are close to each other, one femtocell uplink causes interference to the neighbor FAP receiver (scenarios *C* and *D*). Figure 2 presents an example where the uplink from femto UE-2 causes interference to macrocellular BS receiver and to UE-1 femtocell uplink at FAP-2.

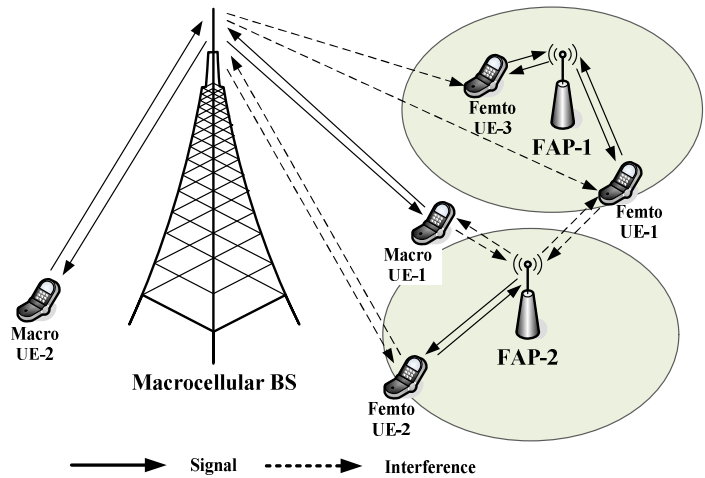

**Fig. 2:** Example of interference scenarios in integrated femtocell/macrocell networks

## 3. The Proposed Frequency Allocation Schemes

In this section, we use the parameters as defined in Table 1 below.

**Table 1:** Basic Nomenclature

| Symbol | Definition |
|---|---|
| $B_T$ | The total system-wide spectrum of frequencies (frequency band) allocated to for macrocells and femtocells |
| $B_m$ | The total frequency spectrum (frequency band) allocated to macrocells |
| $B_f$ | The total frequency spectrum (frequency band) allocated to femtocells |
| $B_{f1}, B_{f2},$ and $B_{f3}$ | The total frequency spectrum (frequency band) allocated to all the femtocells in the macrocells #1, #2, and #3, respectively, of a macrocell cluster |
| $B'_{f1}, B'_{f2},$ and $B'_{f3}$ | The actual frequency band allocation to a femtocells in the macrocell #1, #2, and #3, respectively, of a macrocellular cluster |
| $B_{fnc}$ | The frequency spectrum (frequency band) allocated to the center of a newly installed femtocell |
| $B_{fne}$ | The frequency spectrum (frequency band) allocated to the edge of a newly installed femtocell |
| $B_{foc(k)}$ | The frequency spectrum (frequency band) allocated to the center of a *k-th* overlapping interfering femtocell |
| $B_{foe(k)}$ | The frequency spectrum (frequency band) allocated to the edge of a *k-th* overlapping interfering femtocell |

The cellular spectrum is a quite limited and expensive resource, so spatial reuse of radio spectrum has been a well-known technique for cost reduction. Though, allocation of the same frequencies among neighboring femtocells and overlaid macrocells can potentially cause serious interference effects especially for the dense femtocells deployment case. However, a properly designed frequency allocation scheme can mitigate the interference effects, while improving the utilization of the frequency spectrum. The frequency allocation for the different femtocellular network deployment scenarios should differ to achieve better utilization of the spectrum. In this section, we propose possible efficient and cost effective frequency planning for different femtocellular network deployment scenarios. The total system-wide cellular frequency spectrum, the frequency spectrum allocated for macrocells, and the frequency spectrum allocated to femtocells, are denoted as $B_T$, $B_m$, and $B_f$ respectively.

### 3.1 The *Dedicated Frequency Band* Allocation Scheme

The *dedicated frequency band* allocation is the case where the same frequency band is shared by all the femtocells and a different frequency band is allocated to the macrocells; i.e., in this scheme, femtocells and macrocells use totally separate frequency bands. In the example in Fig. 3, the femtocells use the frequency band from $f_1$ to $f_3$, whereas the frequency band allocation for the macrocells is $f_3$ to $f_2$. This scheme is not suitable to support dense femtocells deployment (scenarios *D*), as the use of the same frequencies by the densely located femtocells of scenario *D* would cause severe interference problem. On the other hand, use of this allocation scheme in the deployment scenarios *A* would cause significant inefficiency. Thus, this scheme could be used only for initial and relatively small-scale deployment of a femtocellular network.

For this scheme, we can write:

$$\left.\begin{matrix} B_m = f_2 - f_3 \\ B_f = f_3 - f_1 \end{matrix}\right\} \quad (1)$$

$$\left.\begin{matrix} B_T = B_m + B_f \\ B_m \cap B_f = \emptyset \\ B_m \cup B_f = B_T \end{matrix}\right\} \quad (2)$$

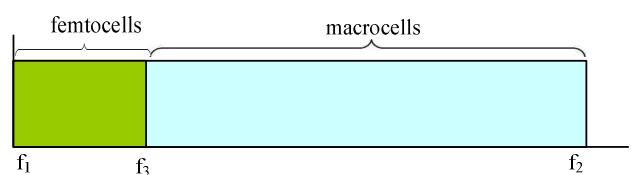

**Fig. 3:** Frequency allocation using the *dedicated frequency band* scheme

### 3.2 *The Shared Frequency Band* Allocation Scheme

In the *shared frequency band* allocation scheme, the frequencies from the same spectrum can be allocated for the femtocells and the macrocells, as shown in Fig. 4. This scheme is very efficient for the deployment scenarios *A* resulting in the best utilization of the spectrum because in this deployment scenario there are no interfering overlapping femtocells and/or overlaying macrocell. This scheme can also be used for the deployment scenario *B* with a small numbers of discrete femtocells overlaid by a macrocell. However, this scheme is inappropriate for other deployment scenarios due to the amount of interference that could be present.

For this scheme, we can write:

$$B_m = B_f = B_T = f_2 - f_1 \quad (3)$$

$$\left.\begin{matrix} B_m \cap B_f = B_T \\ B_m \cup B_f = B_T \end{matrix}\right\} \quad (4)$$

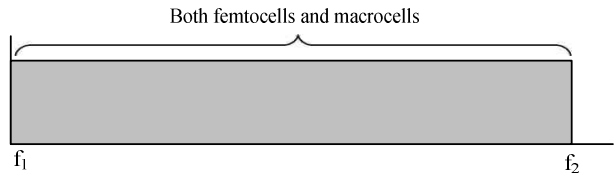
**Fig. 4:** Frequency allocation using the *shared frequency band* scheme

### 3.3 The *Sub-Frequency Band* Scheme

In the *sub-frequency band* allocation scheme, the macrocells use the total system spectrum, while only part of this total frequency band can be used by the femtocells. This is depicted in Fig. 5, where the femtocells use frequency band $f_1$ to $f_4$, while the total frequency allocation for the macrocells is $f_1$ to $f_2$. This scheme cannot support dense femtocells deployment (e.g., deployment scenario *D*) and can also cause significant interference in deployment scenarios *C*. However, in the deployment scenario *B*, the amount of femtocell-to-femtocell interference is limited. This scheme is mostly useful for the deployment scenario *A*. For this scheme, we can write the following set of equations:

$$\left.\begin{array}{r}B_m = B_T = f_2 - f_1 \\ B_f = f_4 - f_1\end{array}\right\} \quad (5)$$

$$\left.\begin{array}{r}B_m \cap B_f = B_f \\ B_m \cup B_f = B_T\end{array}\right\} \quad (6)$$

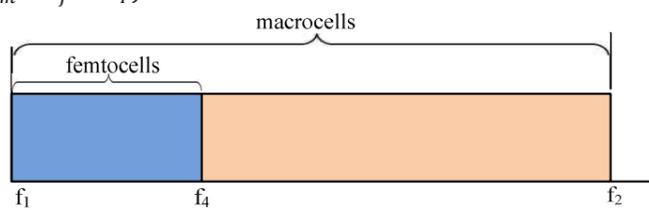
**Fig. 5:** Frequency allocation using *sub-frequency band* scheme

### 3.4 Reuse Frequency among Femtocells and Macrocells

In a large scale deployment of femtocellular networks, the implementation of frequency-reuse can be performed in a number of ways. Here, we present two alternatives for frequency reuse. In both the proposed schemes, we assume that the reuse factor in the macrocellular network is 3 (i.e., the number of macrocells in a macrocellular cluster)..

### 3.4.1 Scheme 1 (the *static frequency-reuse* scheme)

In the *static frequency-reuse* scheme, the set of all cellular frequencies is divided into three equal bands: $B_{m1}$, $B_{m2}$, and $B_{m3}$ and each one of the three macrocells in a macrocellular cluster uses one of these three different frequency bands. If a macrocell uses a particular frequency band, then the femtocells within that macrocell use the other two frequency band. Figs. 6 and 7 show an example of the *static frequency-reuse* scheme. Macrocell #1 of the macrocell cluster uses frequency band $B_{m1}$ and each femtocell within this macrocell uses either $B_{m2}$ or $B_{m3}$. The use of two different frequency bands for neighboring femtocells reduces the femtocell-to-femtocell interference. The deployment scenarios *B* and *C* are quite effective for use with the *static frequency-reuse* scheme. While deployment scenario *D* could also be used with this scheme, but cannot mitigate interference well. SON-based FAPs can be optionally used for automatic assignment of frequencies to femtocells. For this scheme, we can write the following set of equations:

$$\left.\begin{aligned}
B_{m1} &= B_{m2} = B_{m3} = \frac{B_m}{3} \\
B'_{f1} &= a.B_{m2} + a'.B_{m3} \\
B'_{f2} &= b.B_{m3} + b'.B_{m1} \\
B'_{f3} &= c.B_{m1} + c'.B_{m2} \\
B_{f1} &= B_{m2} + B_{m3} \\
B_{f2} &= B_{m3} + B_{m1} \\
B_{f3} &= B_{m1} + B_{m2}
\end{aligned}\right\} \tag{7}$$

$$\left.\begin{aligned}
B_{m1} \cap B_{f1} &= \emptyset,\ B_{m2} \cap B_{f2} = \emptyset,\ B_{m3} \cap B_{f3} = \emptyset \\
B_{m1} \cup B_{f1} &= B_T,\ B_{m2} \cup B_{f2} = B_T,\ B_{m3} \cup B_{f3} = B_T
\end{aligned}\right\}, \tag{8}$$

where *a, b,* and *c* are binary variables and $a'$, $b'$, and $c'$ are their respective complements. $B_{m1}$, $B_{m2}$, and $B_{m3}$ refer the frequency band allocations for macrocell #1, #2, and # of macrocell cluster, respectively. $B_{f1}$, $B_{f2}$, and $B_{f3}$ refer the total frequency band allocations for all the femtocells in the macrocells #1, #2, and #3, respectively, of a macrocellular cluster. $B'_{f1}$, $B'_{f2}$, and $B'_{f3}$ refer to a particular frequency band allocated to a femtocell in the macrocell #1, #2, and #3 of a macrocellular cluster, respectively.

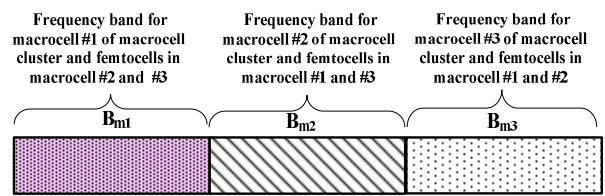

**Fig. 6:** Frequency allocation for macrocells using the *static frequency-reuse* scheme

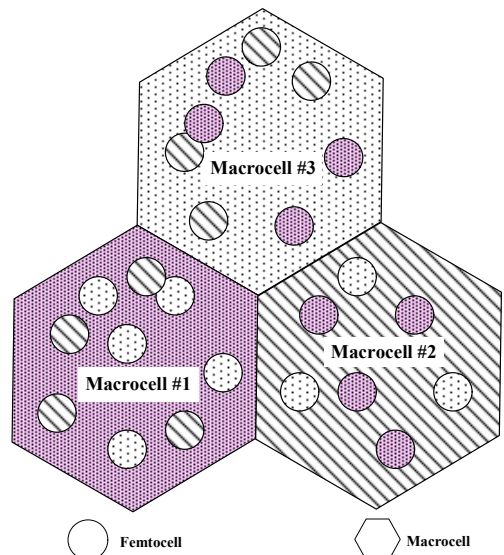

**Fig. 7:** An example of frequency allocation to macrocells of a macrocell cluster and to femtocells within the macrocells for the *static frequency-reuse* scheme

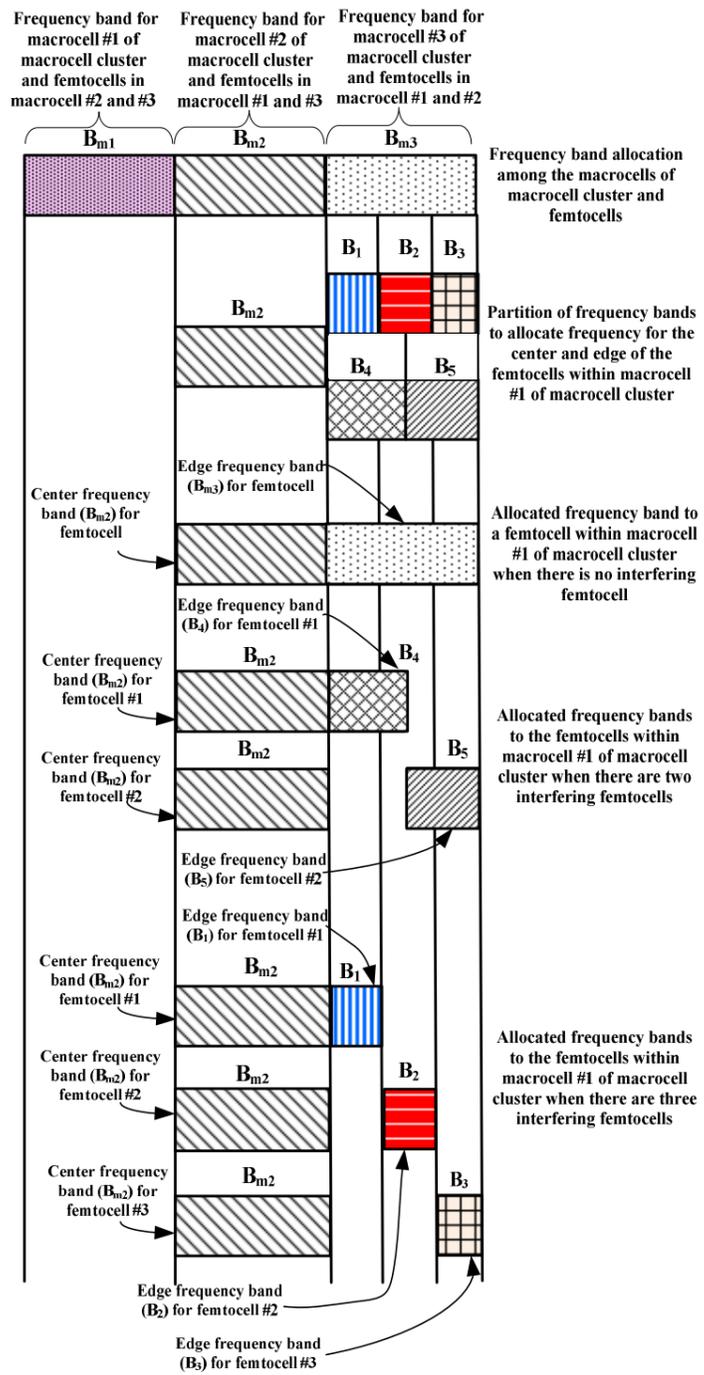

**Fig. 8:** The division of frequency spectrum for the *dynamic frequency-reuse* scheme

### 3.4.2 Scheme 2 (the *dynamic frequency-reuse* scheme)

The total frequency allocation for the three macrocells in a macrocell cluster and the total frequency allocation for the femtocells in each of these three macrocells are similar to the *static frequency-reuse* scheme. However, as opposed to the *static frequency-reuse* scheme where each femtocell uses only one of the two frequency bands which is not assigned to its underlying macrocell, in the *dynamic frequency-reuse* scheme, each femtocell uses two frequency bands. In each femtocells, one band is used in the center of the femtocell, while the other band is used at the edge of the femtocell. The frequency band used in the center of all the femtocells of the same macrocell is the same. However, the frequency bands used in the edges of the various femtocells are, in general, different, to avoid the interference. Figs. 8 and 9 show an example of the frequency allocation for the *dynamic frequency-reuse* scheme. For clarity, only the femtocells of one macrocell (macrocell #1) are shown in the example. The overlapping of femtocells and the interfering signals from other femtocells in this proposed *dynamic frequency-reuse* scheme is mitigated through the use of the different bands at the edges of the femtocells.

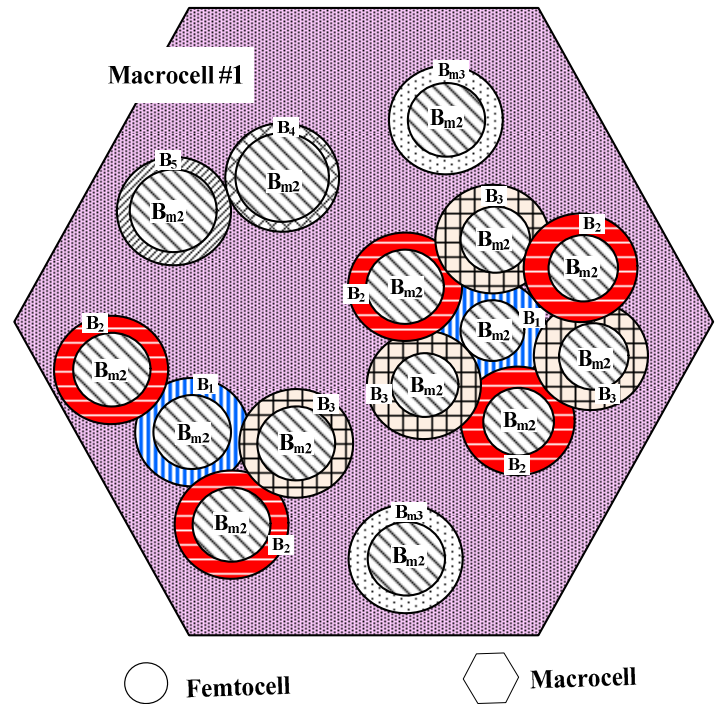

**Fig. 9:** Frequency allocation in the macrocell #1 and the femtocells within this macrocell using the *dynamic frequency-reuse* scheme

For the *dynamic frequency-reuse* scheme, we can write the following set of equations:

$$\left. \begin{aligned} B_{m1} &= B_{m2} = B_{m3} = \frac{B_m}{3} \\ B_1 &= B_2 = B_3 = \frac{B_{m3}}{3} \\ B_4 &= B_5 = \frac{B_{m3}}{2} \\ B_{f1} &= B_{m2} + B_{m3} \\ B_{f2} &= B_{m3} + B_{m1} \\ B_{f3} &= B_{m1} + B_{m2} \end{aligned} \right\} \quad (9)$$

$$B'_{f1} = \begin{cases} B_{m2} + a.B_1 + b.B_2 + c.B_3 & for\ 3\ interfering\ femtocells \\ B_{m2} + x.B_4 + x'.B_5 & for\ 2\ interfering\ femtocells \\ B_{m2} + B_{m3} & for\ no\ interfering\ femtocell \end{cases} \quad (10)$$

$$\left.\begin{array}{l} B_{m1} \cap B_{f1} = \emptyset,\ B_{m2} \cap B_{f2} = \emptyset,\ B_{m3} \cap B_{f3} = \emptyset \\ B_{m1} \cup B_{f1} = B_T,\ B_{m2} \cup B_{f2} = B_T,\ B_{m3} \cup B_{f3} = B_T \end{array}\right\}, \quad (11)$$

where $a$, $b$, $c$ are the binary parameters and $x'$ is the complement of $x$. The value of $a$, $b$, and $c$ depend on the neighboring femtocell's edge frequency band. $B_{m1}$, $B_{m2}$, and $B_{m3}$ refers the frequency band allocation for macrocell #1, #2, and #, respectively, of the macrocell cluster. $B_{f1}$, $B_{f2}$, and $B_{f3}$ refers the total frequency band allocation for all the femtocells in macrocell #1, #2, and #3, respectively, of the macrocell cluster. $B'_{f1}$ refers to the actual frequency band allocation of a femtocells in macrocell #1 (of the macrocell cluster).

**Algorithm to select the frequency band for a newly installed femtocell:**

```
1:  while the detected frequency band of overlaid macrocell = B_m1
2:      the total frequency band allocation for all femtocells in the macrocell = B_m2 + B_m3
3:      if a newly installed femtocell does not detect any interfering femtocell then
4:          B_fnc = B_m2;
5:          B_fne = B_m3;
6:      end if
7:      if a newly installed femtocell detects one interfering femtocell then
8:          B_fnc = B_m2;
9:          if B_foe(1) = B_5 then
10:             B_fne = B_4;
11:         else if B_foe(1) = B_4 then
12:             B_fne = B_5;
13:         else if B_foe(1) = B_1 then
14:             B_fne = B_2;
15:         else if B_foe(1) = B_2 then
16:             B_fne = B_3;
17:         else if B_foe(1) = B_3 then
18:             B_fne = B_1;
19:         end if
20:     end if
21:     if a newly installed femtocell detects two common interfering femtocells then
22:         B_fnc = B_m2;
23:         if B_foe(1) = B_4 and B_foe(2) = B_5 then
24:             B_fne = B_3;
25:             B_foe(1) = B_1;
26:             B_foe(2) = B_2;
27:         else if B_foe(1) = B_1 and B_foe(2) = B_2 then
28:             B_fne = B_3;
29:         else if B_foe(1) = B_2 and B_foe(2) = B_3 then
30:             B_fne = B_1;
31:         else if B_foe(1) = B_3 and B_foe(2) = B_1 then
32:             B_fne = B_2;
33:         end if
34:     end if
```

The radius of the inside circle of a femtocell can vary based on the number of neighbor femtocells and their relative distance. Deployment scenarios *C* and *D* can be quite effectively used with this scheme. Although deployment scenario *B* could also be used with this scheme, however, it would be inefficient. The use of SON-based network feature is essential for the *dynamic frequency-reuse* scheme; e.g., using the SON functionalities, the transmitted power can be automatically adjusted and the edge frequencies can be automatically assigned [8]. The frequency allocation among macrocells in a macrocellular cluster can follow similar procedure. The algorithm to configure the frequency for a newly installed femtocell (consider as an example only one macrocell within a macrocellular cluster of size 3) is proposed below.

Whenever a FAP is removed from a femtocellular network, the existing femtocells re-configure their frequencies in a similar way to the case when a new femtocell is installed. In the proposed scheme, maximum three overlapped femtocells are considered. In a practical case where more than three femtocells overlap, then the size of the femtocells will need to be reduced or re-adjusted automatically to reduce the inter-femtocell interference effect. In a very dense femtocells area, if the best frequency reconfiguration is not possible using the available combination of frequency bands, then the femtocell size will need to be reduced or re-adjusted automatically to mitigate the inter-femtocell interference effect. Thus, the SON-based network architecture is essential for the *dynamic frequency-reuse* scheme in the femtocellular network environment.

## 4. The Proposed SON Architecture for Highly Dense Deployment

As mentioned in the previous section, the SON architecture is required to support the *dynamic frequency-reuse* scheme. The main functionalities of the SON architecture are: self-configuration, self-optimization, and self-healing. The self-configuration feature allows intelligent frequency allocation among neighboring FAPs, maintenance of the neighboring cell list, and support for mobility. The self-optimization feature optimizes the settings of transmission powers of neighboring FAPs, as well as other operational parameters, such as the size of the neighbor list. The self-healing feature supports automatic detection and resolution of major failures. A "*sniffing*" function is required for effective integration of femtocells into a macrocellular network, so that a FAP can scan the air interface and detect available frequencies and other network resources. Communication among neighboring FAPs, as well as between FAPs and the respective macrocellular BS, is required to configure spectral resources and transmission powers. Therefore, further augmentation of the SON architecture would be imperative for a full scale femtocellular network deployment. Fig. 10 shows the framework, and the basic features, of an integrated SON femtocell/macrocell architecture. Next, we discuss three representative cases of operation of a femtocellular network enhanced with SON features.

- *Case 1 (Frequency configuration among neighboring femtocells):* When a FAP is newly installed, it configures its center and edge frequencies according to the detected frequencies of the neighboring femtocells. The FAPs of the entire neighborhood coordinate this frequency allocation.

- *Case 2 (Cell size re-adjustment):* If a number of femtocells interfere with each other even after the frequency configuration procedure was performed, the affected FAPs coordinate among themselves to re-adjust their center and edge areas, as to reduce the interference effect.

- *Case 3 (Frequency configuration between a newly installed FAP and macrocellular BS):* When a FAP is newly installed in a macrocellular coverage area, the FAP and the respective macrocellular BS communicate with each other to configure the frequencies of the newly installed FAP.

In addition to frequency allocation schemes, there are other techniques that are also used to mitigate interference in a SON-based femtocellular network. Power optimization technique can be quite effective in areas of dense femtocell deployment and for femtocells located on the fringe of a macrocell. Based on the presence of femto users and their operating modes, the FAP can change its mode of operation to mitigate the interference among the neighboring femtocells. For instance, the users on the edge of a macrocell typically transmit higher power that causes interference to the neighboring FAPs. A FAP may increase its cell size to accept such macro users into its femtocell. Thus by adjusting the cell size, the interference to the neighboring femtocells can be reduced. As another example, power control of the UE is needed when the distance between a FAP and the respective macrocellular BS is short.

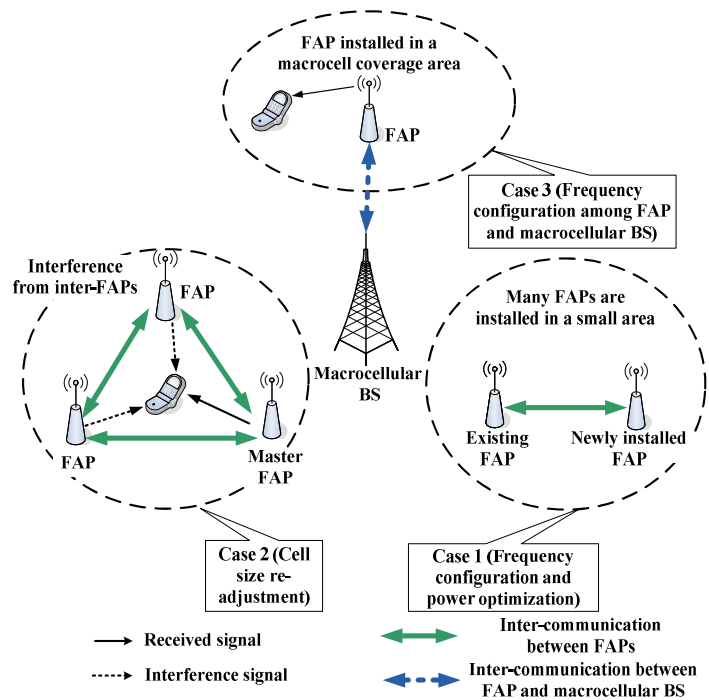

**Fig. 10:** The SON features for a highly dense deployment of an integrated femtocell/macrocell network

## 5 Outage Probability Analyses

For the analysis in this section, we use the parameters as defined in Table 2.

**Table 2:** Nomenclature for Outage Probability Analysis

| Symbol | Definition |
|---|---|
| $S_o$ | The received signal from the associated (reference) FAP |
| $I_{m(j)}$ | The received interference from the *j-th* interfering macrocellular BS |
| $K$ | The total number of neighboring femtocells |
| $N$ | The total number of interfering macrocells |
| $I_{n(i)}$ | The received interference from the *i-th* neighboring femtocell |

| $I_f$ | The total received interference from all the neighboring femtocells |
| --- | --- |
| $P_T$ | The transmitted signal power |
| $P_R$ | The signal power received at the receiver |
| $d$ | The distance between the transmitter and the receiver |
| $\xi$ | The slow-fading random variable |
| $Z$ | The fast-fading random variable |
| $\eta$ | The path loss exponent |

There are various interference mechanisms present in the macrocell/femtocell integrated network architecture; in particular, between macrocells and femtocells, and among famtocells. Inadequate interference management system reduces system capacity for both the macrocellular and the femtocellular networks. More specifically, interference reduces users' QoE, and causes higher outage probability. On the other hand, appropriate interference management can increase the capacities of both, the femtocellular and the macrocellular networks.

Assuming that the spectrums of the transmitted signals are spread, we can approximate the interference as AWGN. Then, following the Shannon Capacity Formula,

$$C = W \log_2(1 + SNR) \text{ [bits/sec]},$$

we can state that the capacity of a wireless channel decreases with decreased signal-to-interference (*SIR*) level. The received *SIR* level of a femtocell user in a macrocell/femtocell integrated network can be expressed as:

$$SIR = \frac{S_o}{\sum_{j=0}^{N-1} I_{m(j)} + \sum_{i=1}^{K} I_{n(i)}}, \qquad (12)$$

where $S_o$ is the power of the signal from the associated (reference) femtocell, $I_{m(j)}$ is the power of the interference signal from the *j-th* interfering macrocell from among the *N* interfering macrocells, and $I_{n(i)}$ is the received interference signal from the *i-th* femtocell from among the *K* neighboring femtocells. The indices *0*, *i*, and *j* refer to the reference femtocell, the *i-th* neighboring femtocell, and the *j-th* interfering macrocell, respectively.

The outage probability of a femtocell user can be calculated as:

$$P_{out} = P_r(SIR < \gamma), \qquad (13)$$

where *γ* is a threshold value of *SIR* below which there is no acceptable reception. Alternatively, equation (13) can be rewritten as:

$$\begin{aligned} P_{out} &= P_r \left( \frac{S_o}{\sum_{j=0}^{N-1} I_{m(j)} + \sum_{i=1}^{K} I_{n(i)}} < \gamma \right) \\ &= P_r \left\{ \sum_{i=1}^{K} I_{n(i)} > \left( \frac{S_o}{\gamma} - \sum_{j=0}^{N-1} I_{m(j)} \right) \right\} \\ &= P_r \left\{ I_f > \left( \frac{S_o}{\gamma} - I_m \right) \right\}, \end{aligned} \qquad (14)$$

where $I_f$ and $I_m$ are the total received interference from the neighbor femtocells and macrocells, respectively. We assume that the probability density function (PDF) of $I_f$ is Gaussian [10].

As per equation (14), to reduce the probability of outage, techniques to mitigate the interference from the interfering macrocells and the neighboring femtocells need to be employed. The choice of such a technique varies according to the macrocell coverage area and the density of the femtocells. For example, in remote areas, where macrocell coverage is sparse or not available at all, or when the macrocell and the femtocells use non-overlapping frequency bands, the interference from macrocells can be assumed negligible and $\forall_j$, $I_{m(j)} \approx 0$. In a dense femtocellular deployment, the interference form neighboring femtocells is of main concern and proper inter-femtocell interference management can suffice to increase the signal-to-interference ratio and to reduce the outage probability. Fig. 11 depicts the various signals and interfering signals affecting the outage probability of a femtocell user.

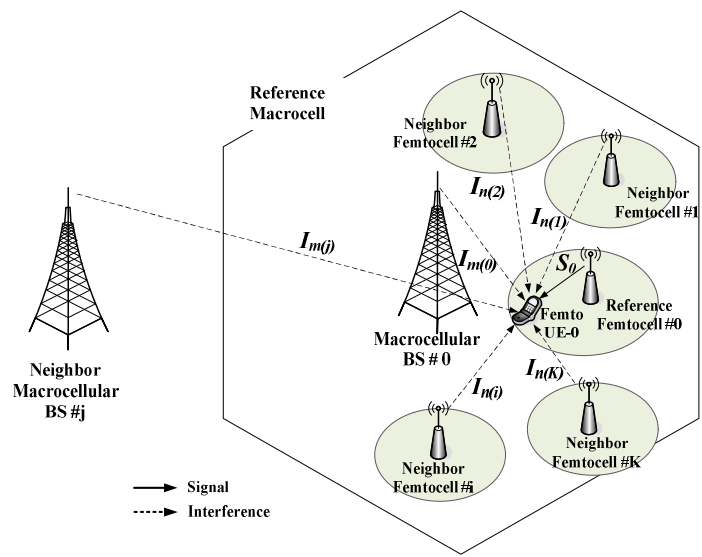

**Fig. 11:** Signals and interferences for a user situated in the reference femtocell

The received signal power $P_R$ from a transmitter located at distance $d$ from the receiver, when the transmitted signal is of power $P_T$ is:

$$P_R = P_T P_0 \, d^{-\eta} \xi Z \, , \tag{15}$$

where $P_0$ is a function of the carrier frequency, the antenna height, and the antenna gain; $\xi$ is a random variable that accounts for slow-fading (the so-called "shadowing") of the radio channel; $Z$ is a random variable that represents the effect of fast fading (i.e., multi-path effect) of the communication channel; and, finally, $\eta$ is the path loss exponent. The distribution of $\xi$ is typically assumed to be log-normal [11]. We assume that the envelope of a fast-faded channel is Rayleigh [11] and, therefore, the distribution of the power attenuation factor, $Z$, due to Rayleigh fading is modeled as exponential [12].

When the femtocell user receives signal from its FAP that is situated indoors, then typically the shadowing is negligible and the slow-fading can be neglected. Thus, using equation (15), the received signal by a femtocell user from its associated FAP, $S_o$, cab ne expressed by:

$$S_o = P_{Rf(0)} = P_{Tf(0)} P_{0f} d_0^{-\eta_1} Z_0 = \bar{S} \, Z_0 \tag{16}$$

while the interference $I_{n(i)}$ and $I_{m(j)}$ from the *i-th* neighbor femtocells and *j-th* interfering macrocell, respectively, can be expressed as:

$$I_{n(i)} = P_{Rf(i)} = P_{Tf(i)} P_{0f} d_i^{-\eta_2} \xi_i Z_i \tag{17}$$

$$I_{m(j)} = P_{Rm(j)} = P_{Tm} P_{0m} d_j^{-\eta_3} \xi_{m(j)} Z_{m(j)}, \tag{18}$$

where $i = 0$ and $j = 0$ denote the reference (associated) femtocell and the reference (underlying) macrocell, respectively. $P_{Rm(j)}$ and $P_{Rf(i)}$ stand for the received power at UE from the *j-th* macrocell and the *i-th* femtocell, respectively. $P_{Tm}$ and $P_{Tf(i)}$ represent the transmitted power from each of the macrocells and the *i-th* femtocell respectively. $d_j$ and $d_i$ are the distances from the reference UE to the *j-th* macrocellular BS and to the *i-th* FAP respectively.

We denote by $I_f$ and $I_m$ the power of the total interference from the *K* femtocell neighbors and the power of the total interference from the *N* interfering macrocells BSs, respectively. Using equations (17) and (18), $I_f$ and $I_m$ are derived as:

$$I_f = \sum_{i=1}^{K} I_{n(i)} = \sum_{i=1}^{K} P_{Tf(i)} P_{0f} d_i^{-\eta_2} \xi_i Z_i X_i \tag{19}$$

$$I_m = \sum_{j=0}^{N-1} I_{m(j)} = \sum_{j=0}^{N-1} P_{Tm} P_{0m} d_j^{-\eta_3} \xi_{m(j)} Z_{m(j)} Y_j \tag{20}$$

where $X_i$ and $Y_j$ are binary indication functions which take the value of 1 when the reference femtocell and the *i-th* neighboring femtocell ($X_i = 1$) or the *j-th* macrocell ($Y_j = 1$) use same frequency. Otherwise, $X_i = 0$ or $Y_j = 0$.

Using equations (14) and (16), $P_{out}$ can be written as:

$$P_{out} = P_r \left\{ Z_0 < \frac{\gamma}{S} (I_f + I_m) \right\}. \tag{21}$$

The PDF of $Z_0$ is exponentially distributed. Thus, the solution of equation (21) can be computed as:

$$\begin{aligned} P_{out} &= \int_0^{\frac{\gamma}{S}(I_f + I_m)} \exp(-Z_0) \, dZ_0 \\ &= 1 - \exp\left\{ -\frac{\gamma}{S}(I_f + I_m) \right\} \\ &= 1 - \exp\left[ -\frac{\gamma}{S} \left\{ \sum_{i=1}^{K} I_{n(i)} + \sum_{j=0}^{N-1} I_{m(j)} \right\} \right] \\ &= 1 - \left\{ \prod_{i=1}^{K} e^{\left\{-\frac{\gamma}{S} I_{n(i)} X_i\right\}} \right\} \left\{ \prod_{j=0}^{N-1} e^{\left\{-\frac{\gamma}{S} I_{m(j)} Y_j\right\}} \right\}. \end{aligned} \tag{22}$$

We used equations (16) − (21) in equation (22) to calculate $P_{out}$. In the equation (22), $Y_j = 1$ if the *j-th* macrocell and the reference femtocell are allocated the same frequency, otherwise $Y_j = 0$ (in other words, $B_m \cap B_f = \emptyset$). Thus, for the *shared frequency band* and the *sub-frequency band* schemes, $Y_j = 1$. For the *frequency-reuse* schemes introduced in this paper, $Y_0 = 0$ and for some macrocells in the first and the higher tiers $Y_j = 1$ and for the others $Y_j = 0$. However, the macrocells in the second and in the higher tiers cause only negligible interference. The value of $X_i$ is related to the allocated frequency for the reference femtocell and the *i-th* neighbor femtocell. $X_i = 1$ for the *shared frequency band*, the *dedicated frequency band*, and the *sub-frequency band* schemes, where all the *K* neighboring femtocells use the same frequency as the reference femtocell, For the *static frequency-reuse* scheme, $X_i = 1$ for almost 50% of the neighboring femtocells and $X_i = 0$ for the remaining neighboring femtocells. Similarly, for the *dynamic frequency-reuse* scheme, $X_i = 0$ for more than 66% of the neighboring femtocells. Thus, the outage probability for the *dynamic frequency- reuse* scheme is lower than other schemes.

## 6. Numerical Results

In this section, we evaluate the throughput and the outage probability of the proposed *static* and *dynamic frequency- reuse* schemes and compare with the *shared frequency band* scheme and the *dedicated frequency band* scheme for various femtocells densities.

In our evaluation, we define as dense femtocells deployment (scenario *D*) as more than 1000 femtocells within a macrocell; otherwise, we consider it as non-dense femtocellular deployment (scenarios *B* and *C*). Table 2 summarizes the values of the parameters that we used in our numerical analysis. For simplicity, we used the macrocell propagation model from [9] and the femtocell propagation model from [2]. We neglect the effects of the macrocells in the second and the higher tiers, as the contributed interference is minimal. Only the reference macrocell and the other six macrocells in the first tier are considered. Furthermore, for the purpose of calculation of inter-femtocell interference, we assume that there exists one wall between two femtocells. We consider two femtocells as neighbors, if their FAPs are within 60 meter of each other. The femtocells are placed randomly within the macrocell coverage area and the number of femtocells within the neighbor area is randomly generated according to the Poisson distribution. For the *dedicated frequency band* scheme, we assume 33.3% of total cellular frequency band is allocated to femtocells and the remaining 66.7% of the total frequency band is allocated to the macrocells. For capacity analysis we use the Shannon Capacity Formula.

**Table 3:** Summary of the parameter values used in our analysis

| Parameter | Value |
|---|---|
| Macrocell radius | 1 [km] |
| Femtocell radius | 10 [m] |
| Distance between the reference macrocellular BS and the reference FAP | 200 [m] |
| Carrier frequency | 900 [MHz] |
| Transmit signal power by the macrocellular BS | 1.5 [kW] |
| Maximum transmitted signal power by a FAP | 10 [mW] |
| Height of a macrocellular BS | 50 [m] |
| Height of a FAP | 2 [m] |
| Threshold value of SIR ($\gamma$) | 9 [dB] |

Fig. 12 demonstrates that the proposed *static frequency-reuse* scheme provides better throughput than the *dedicated frequency band* and the *shared frequency band* schemes for non-dense femtocellular networks deployment. The throughput of the *dynamic frequency-reuse* scheme and of the *static frequency-reuse* scheme are almost same. However, *dynamic frequency-reuse* scheme is inferior for non-dense femtocells deployment, because of the increased implementation cost of the SON features that are needed for the *dynamic frequency-reuse* scheme. Fig. 13 shows that the *static frequency-reuse*-scheme also reduces the outage probability within the considered range of parameters. The outage probability of the *dynamic frequency-reuse* scheme is almost same as that of the *static frequency-reuse* scheme for non-dense femtocellular network deployment. Thus, the *static frequency-reuse* scheme is recommended for the non-dense femtocellular network deployment.

Fig. 14 depicts the throughput performance of the *dynamic frequency-reuse* scheme in dense femtocells environment. The throughput of the *dynamic frequency-reuse* scheme is larger than that of the other schemes for dense femtocellular deployment. However, the throughput quickly degrades to small values for the *shared frequency band* scheme and for the *dedicated frequency band* scheme. Fig. 15 illustrates the fact that that the outage probability of the *dynamic frequency-reuse* scheme is significantly smaller compared with the other schemes even for highly dense femtocellular network deployment. The results in Figs. 14 and 15 were obtained when the distance between the femto UE and the reference FAP was maintained at 5 meters.

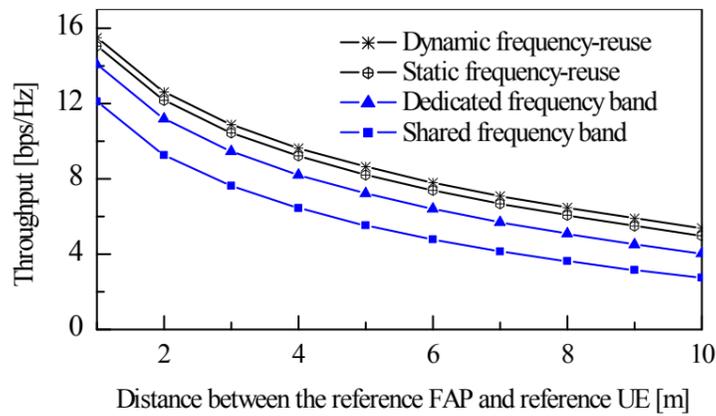

**Fig. 12:** Throughput comparison of non-dense femtocellular network deployment (scenarios *B* and *C*)

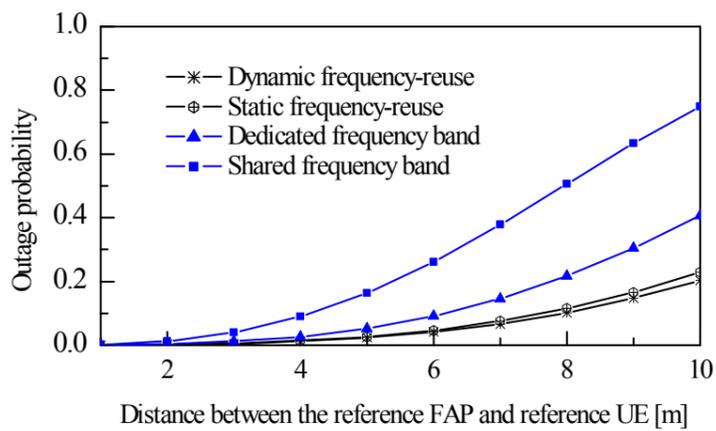

**Fig. 13:** Outage probability comparison of non-dense femtocellular network deployment (scenarios *B* and *C*)

### 6.1 Summary of Numerical Results

The *dynamic frequency-reuse* scheme outperforms the other schemes investigated in this paper for all the femtocellular network environments in terms of throughput, outage probability, and spectral use. However, due to the overhead associated with the implementation of the SON features, the *dynamic frequency-reuse* scheme is not appropriate for small-scale femtocellular network deployment. On the other hand, SON-based network architecture is not essential for the *static frequency-reuse* scheme. Therefore, since in the non-dense femtocellular network deployment the performance of the *static frequency-reuse* scheme and the *dynamic frequency-reuse* scheme are essentially equal, the *static frequency-reuse* scheme is preferable for the non-dense deployment cases (scenarios *B* and *C*).

For the scenario *A*, the *shared frequency band* should be used to increase the spectral efficiency.

We recommend that for the dense femtocellular network deployment (scenario *D*) the use of *dynamic frequency-reuse* scheme should be adopted.

Finally, we point out that if the implementation of the SON features in the femtocellular network architecture is not expensive or if cost is not a major factor (as may be the case in some military installations), then the *dynamic frequency-reuse* scheme is, indeed, the best choice for all types of femtocellular network deployments.

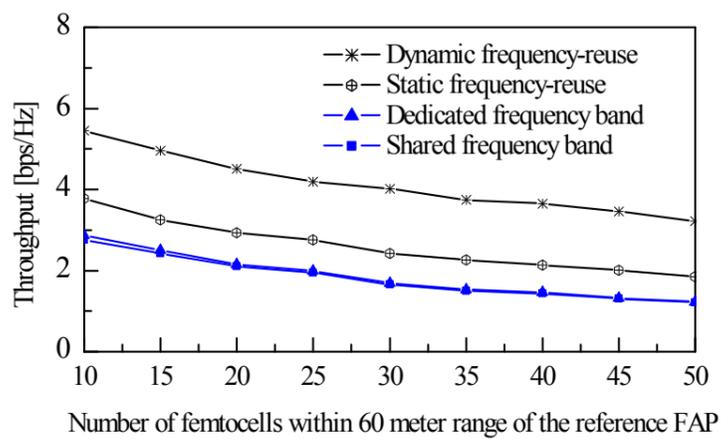

**Fig. 14:** Throughput comparison of the dense femtocells scenario

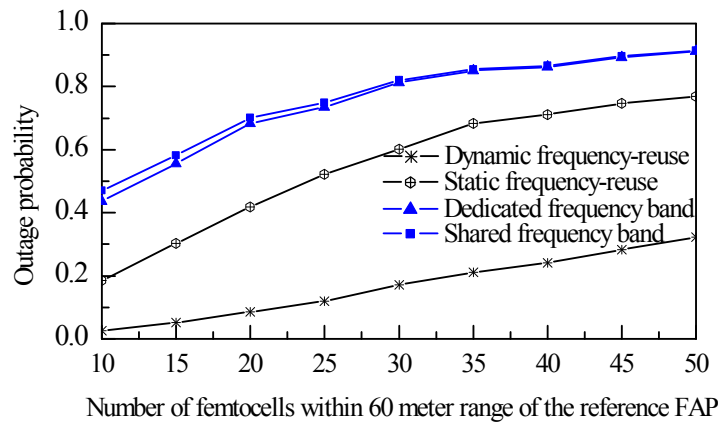

**Fig. 15:** Outage probability of the dense femtocells scenario

## 7. Conclusions

Femtocells are a novel wireless networking technology that holds the potential of increasing the cellular network capacity by offloading some of the traffic of the macrocellular installations. However, the main advantage of the femtocellular technology – the use of the same radio spectrum as that of the macrocellular systems – becomes also the source of considerable challenge in implementation of the femtocellular networks. This is so, as due to the sharing of the radio spectrum, there is a possibility of substantial interference between the macro- and the femto-networks. This is, in addition to the interference among the femtocells themselves. Consequently, the channel (frequency) allocation problem needs to be satisfactorily addressed, as improper frequency management for the integrated femtocell/macrocell networks may cause severe interference problems as to even totally eliminate the advantage of the femtocellular technology altogether.

Femtocellular networks may come in different sizes and ultimately we expect to see densely deployed networks with over thousand of femtocells overlaid by a single macrocell. To be effective, different approaches should be taken for different scenarios of femtocellular network deployment. In this paper, we classified the different deployment scenarios based on the numbers of femtocells within a macrocell and their locations relative to one another. We then studied a number of frequency allocation schemes for two main cases: dense and non-dense femtocell deployment, where the threshold that differenciate the two cases is 1000 femtocells per macrocell. Furthermore, we include in our consideration the possibility of implementation of the SON features, which would be quite effective in re-configuration of frequency among neighbor femtocells and between femtocells and macrocells, in re-adjusting the cell size, and in optimization of the transmission powers.

Our numerical results demonstrate that different schemes are preferable for different femtocellular network deployment to mitigate the interference; to increase the throughput; to reduce the outage probability; and to increase the spectral efficiency. In general, the *dynamic frequency-reuse* scheme outperforms, or is similar to, the other schemes. However, this scheme also requires implementation of the SON features, which might prove to be counterproductive for the small gain of the scheme in non-dense femtocellular networks. Thus, we advocate the use of the *dynamic frequency-reuse* scheme for dense deployment only, unless the cost of the SON implementation is either small or not a major factor.

Our study shed some light onto the challenging problem of interference mitigation in an integrated femtocellular/macrocellular networks. Nevertheless, much work remains to be done to address this critical problem, which might be the key to proliferation of this technology on a broad scale.

## Acknowledgement

This work was supported by the IT R&D program of MKE/KEIT [10035362, Development of Home Network Technology based on LED-ID]. The work of Zygmunt Haas was supported by the grant from the US National Science Foundation number CNS-0626751 and by the US AFOSR contract number FA9550-09-10121.

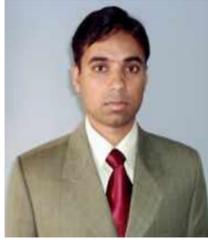

**Mostafa Zaman Chowdhury** received his B.Sc. in 2002 Electrical and Electronic Engineering from Khulna University of Engineering and Technology (KUET), Bangladesh. In 2003, he joined the Electrical and Electronic Engineering department of KUET, Bangladesh as a faculty member. He received his M.Sc. in Electronics Engineering from Kookmin University, Korea in 2008. Currently he is working towards his Ph.D. degree in the department of Electronics Engineering at the Kookmin University, Korea. He served as a reviewer for several international journals (including IEEE Communications Magazine and IEEE Communications Letters) and conferences (including IEEE conferences). He has been involved in several Korean Government projects. His current research interests focus on convergence networks, QoS provisioning, mobility management, and femtocell networks.

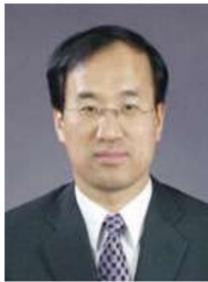

**Yeong Min Jang** received the B.E. and M.E. degree in Electronics Engineering from Kyungpook National University, Korea, in 1985 and 1987, respectively. He received the doctoral degree in Computer Science from the University of Massachusetts, USA, in 1999. He worked for ETRI between 1987 and 2000. Since September 2002, he is with the School of Electrical Engineering, Kookmin University, Seoul, Korea. He has organized several conferences such as ICUFN2009 and ICUFN2010. He is currently a member of the IEEE and KICS (Korea Information and Communications Society). He received the Young Science Award from the Korean Government (2003 to 2006). He had been the director of the Ubiquitous IT Convergence Center at Kookmin University since 2005. He has served as the executive director of KICS since 2006. He has served as a founding chair of the KICS Technical Committee on Communication Networks in 2007 and 2008. His research interests include IMT-advanced, radio resource management, femtocell networks, Multi-screen convergence networks, and VLC WPANs.

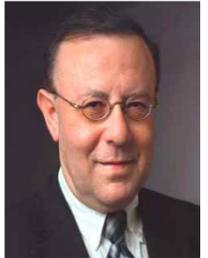

**Zygmunt J. Haas** received his B.Sc. in ECE in 1979, his M.Sc. in EE in 1985, and his Ph.D. from Stanford University in 1988. Subsequently, he joined the AT&T Bell Laboratories in the Network Research Department. There he pursued research on wireless communications, mobility management, fast protocols, optical networks, and optical switching. From September 1994 till July 1995, Dr. Haas worked for the AT&T Wireless Center of Excellence, where he investigated various aspects of wireless and mobile networking, concentrating on TCP/IP networks. In August 1995, he joined the faculty of the School of Electrical and Computer Engineering at Cornell University. He directs the *Wireless Networks Laboratory (WNL)*, an internationally recognized research group specializing in ad hoc and sensor networks.

Dr. Haas is an author of numerous technical papers and holds eighteen patents in the fields of high-speed networking, wireless networks, and optical switching. He has organized several workshops, delivered numerous tutorials at major IEEE and ACM conferences, and has served as editor of a number of journals and magazines, including the IEEE Transactions on Networking, the IEEE Transactions on Wireless Communications, the IEEE Communications Magazine, and the Springer "Wireless Networks" journal. He has also been the guest editor of several IEEE JSAC issues. Dr. Haas served in the past as a Chair of the IEEE Technical Committee on Personal Communications (TCPC). He is an IEEE Fellow. His interests include: mobile and wireless communication and networks, biologically-inspired networks, and modeling of complex systems.